\documentclass{article}
\usepackage{spconf,amsmath,graphicx}

\usepackage{enumitem}

\usepackage{mathtools}

\usepackage{siunitx}

\usepackage[acronym]{glossaries}
\usepackage{balance}

\usepackage{hyperref}
\usepackage{cleveref}
\usepackage{tabularx, etoolbox, booktabs}
\usepackage{multirow} 
\newcolumntype{H}{>{\setbox0=\hbox\bgroup}c<{\egroup}@{}}  

\usepackage{algorithm,cite}
\usepackage{algpseudocode}

\algtext*{EndWhile}  
\algtext*{EndIf}  
\algtext*{EndFor}

\usepackage{amsmath}
\usepackage{amssymb}
\usepackage{bbm}
\usepackage{physics}
\usepackage{siunitx}
\usepackage{trfsigns}

\newcommand{\vect}[1]{\ensuremath{\boldsymbol{\mathbf{#1}}}}

\def\y{\vect{y}_t}
\def\x{\vect{x}_t}
\def\m{\vect{m}_t}
\def\a{\vect{a}_t}
\def\xh{\hat{\vect{x}}_t}
\def\e{\vect{e}}
\def\Ta{T^a}
\def\p{p_t}

\renewenvironment{thebibliography}[1]{
  \begin{oldthebibliography}{#1}
  \setlength{\itemsep}{0.em}
  \setlength{\parskip}{0.em}
}
{
  \end{oldthebibliography}
}

\ninept

\usepackage[nodisplayskipstretch]{setspace}

\title{Speaker activity driven neural speech extraction}
\name{\begin{tabular}{c}%
Marc Delcroix$^{1}$, %
Katerina Zmolikova$^{2}$\sthanks{Katerina Zmolikova was partly supported by Czech Ministry of Education, Youth and Sports from the National Programme of Sustainability (NPU II) project "IT4Innovations excellence in science - LQ1602"},
Tsubasa Ochiai$^{1}$, %
Keisuke Kinoshita$^{1}$, Tomohiro Nakatani$^{1}$
\end{tabular}}
\address{%
$^{1}$ NTT Corporation, Japan, \\
$^{2}$ Brno University of Technology, Speech@FIT and IT4I Center of Excellence, Czechia 
}
\begin{document}

%
\maketitle
\begin{abstract}
Target speech extraction, which extracts the speech of a target speaker in a mixture given auxiliary speaker clues, has recently received increased interest. Various clues have been investigated such as pre-recorded enrollment utterances, direction information, or video of the target speaker. In this paper, we explore the use of speaker activity information as an auxiliary clue for single-channel neural network-based speech extraction. 
We propose a speaker activity driven speech extraction neural network (ADEnet) and show that it can achieve performance levels competitive with enrollment-based approaches, without the need for pre-recordings.
We further demonstrate the potential of the proposed approach for processing meeting-like recordings, where the speaker activity is obtained from a diarization system. We show that this simple yet practical approach can successfully extract speakers after diarization, which results in improved ASR performance, especially in high overlapping conditions, with a relative word error rate reduction of up to 25~\%.
\end{abstract}
\section{Introduction}
\label{sec:intro}
Recognizing speech in the presence of interfering speakers remains one of the challenges for automatic speech recognition (ASR).
A conventional approach to tackle this problem consists of separating the observed speech mixture into all of its source speech signals before ASR.
Single-channel speech separation has greatly progressed with the introduction of deep learning~\cite{Kolbaek_taslp17,Luo2018tasnet}. However, most separation approaches suffer from two limitations, (1) they require knowing or estimating the number of speakers in the mixture, and (2) they suffer from a global permutation ambiguity issue, i.e., an arbitrary mapping between source speakers and outputs. 

Target speech extraction~\cite{zmolikova2017spkaware} has been proposed as an alternative to speech separation to alleviate the above limitations. It focuses on extracting only the speech signal of a speaker of interest or target speaker by exploiting auxiliary clues about that speaker. The problem formulation becomes thus independent of the number of speakers in the mixture. Besides, the global permutation ambiguity is naturally solved thanks to the use of auxiliary clues.
Several target speech extraction schemes have been proposed, exploiting different types of auxiliary clues such as pre-recorded enrollment utterances of the target speaker~\cite{zmolikova2017learning, Chen2018DeepEN}, direction information~\cite{Gu2019}, video of the target speaker~\cite{ephrat2018looking,afouras2018conversation}, or electroencephalogram (EEG) signals~\cite{CEOLINI2020117282}.  

For example, SpeakerBeam~\cite{zmolikova2017learning,zmolikova2019Journal} is an early approach for enrollment utterance-based target speech extraction. It exploits a speaker embedding vector derived from an enrollment utterance to inform a speech extraction network which speaker to extract from the mixture. With SpeakerBeam, the speaker embedding vectors are obtained from an auxiliary network that is jointly trained with the speech extraction network. Consequently, the speaker embedding vector may capture the voice characteristics of the target speaker, which are needed to distinguish that speaker in a mixture.

Alternative approaches that exploit video clues have been proposed subsequently~\cite{ephrat2018looking,afouras2018conversation}. In these works, a sequence of face embedding vectors are extracted from the video of the face of the target speaker speaking in the mixture and passed to the extraction network. Different from SpeakerBeam, the auxiliary clue is time-synchronized with the mixture signal. 
What information the face embedding vectors actually capture remains unclear, but we can naturally assume that they capture the lip movements of the target speaker and that the mouth opening and closing regions may be important information provided to the speech extraction network.

Although target speech extraction can achieve a high level of performance~\cite{delcroix2020improving,ephrat2018looking}, it is not always possible to have access to the enrollment utterance or video. In this paper, inspired by the works on visual speech extraction, we investigate the use of another clue, which consists of the speech activity of a speaker. The speaker activity consists of a signal that indicates for each time frame if the speaker is speaking or not. It can be a general and practical clue for speech extraction as there are various ways to obtain it.

We propose a single-channel speaker activity driven speech extraction neural network (ADEnet).
We hypothesize that a neural network can exploit the speaker activity information to identify and extract the speech of a speaker, assuming that the speakers in the mixture do not always fully overlap. 
We experimentally demonstrate that assuming the availability of oracle speaker activity, ADEnet can achieve superior performance to SpeakerBeam, without requiring pre-recorded enrollment speech.

In practice, there are various possible approaches to obtain the speaker activity information including visual-based voice activity detection (VAD)~\cite{liu2004voice}, personal VAD~\cite{Ding2020} or diarization~\cite{GarciaRomero2017SpeakerDU,Huang2020SpeakerDW,Medennikov2020TargetSpeakerVA}. For example, speaker diarization has greatly progressed~\cite{Medennikov2020TargetSpeakerVA,GarciaRomero2017SpeakerDU,Huang2020SpeakerDW,Fujita2020EndtoEndND}, and finding speaker activity regions in meetings has achieved a low diarization error rate (DER) even in overlapping conditions. It does not require pre-recorded enrollment utterances of the speakers or video. Diarization is often used as a pre-processing for ASR. However, diarization itself does not extract the speech signals, which limits ASR performance in overlapping conditions. As a practical potential use-case of ADEnet, we investigate using the speaker activity information obtained from a diarization system as clues for ADEnet. We show that this simple approach can improve ASR performance in meeting-like single-channel recordings, especially in severe overlapping conditions. 
Note that our proposed ADEnet does not depend on how the speaker activity information is obtained, making it a versatile speech extraction method. 

\begin{figure*}[tb]
    \centering
    \includegraphics[width=0.98\textwidth]{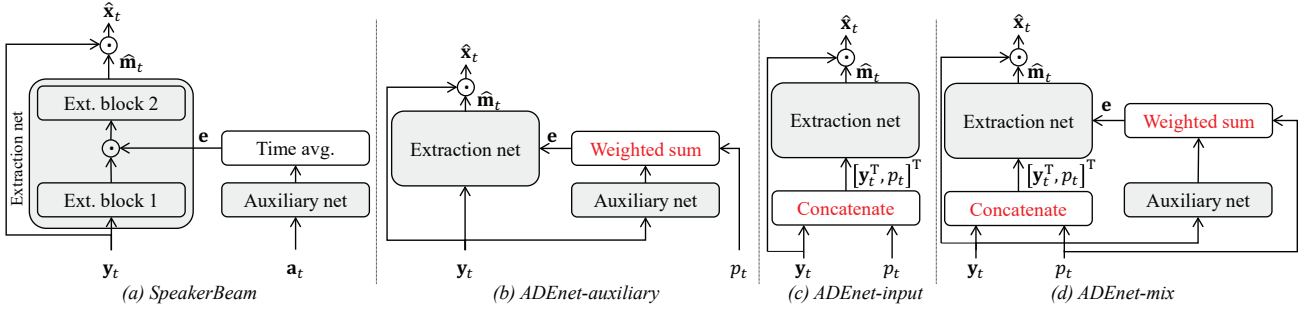}
    \vspace{-4mm}
    \caption{Diagram of SpeakerBeam and  different variations of ADEnet. See Sect. \ref{ssec:settings} for details of the auxiliary and extraction blocks.}
    \vspace{-6mm}
    \label{fig:methods}
\end{figure*}

In the remainder of the paper, we first briefly review SpeakerBeam in Section~\ref{sec:spkbeam}, which serves as a basis for the proposed ADEnet that we describe in Section~\ref{sec:ADEnet}. We discuss related works in Section~\ref{sec:related}. In Section~\ref{sec:expe}, we present experimental results based on the LibriCSS corpus~\cite{Chen2020ContinuousSS}. Finally, we conclude the paper in Section~\ref{sec:conclusion}.

\section{SpeakerBeam}

\label{sec:spkbeam}
Figure \ref{fig:methods}-(a) shows a schematic diagram of SpeakerBeam~\cite{zmolikova2019Journal}. Let $\y$ be an observed speech mixture, $\a$ be an enrollment utterance, and $\x$ be the speech signal of the target speaker.
Here, we consider networks operating in the frequency domain, and the signals  $\y$,  $\x$, and  $\a$ are sequences of amplitude spectrum coefficients, and $t$ represents the time-frame index. Note that we could derive the methods presented in this paper equivalently for time-domain networks~\cite{delcroix2020improving}.

SpeakerBeam is composed of two networks.
First, an auxiliary network accepts the enrollment utterance and outputs a fixed dimension speaker embedding vector $\mathbf{e}$ obtained as the time-average of the output of the auxiliary network,
\begin{equation}
\setlength{\abovedisplayskip}{1pt}
\setlength{\belowdisplayskip}{1pt}
    \e = \frac{1}{\Ta} \sum_{t=1}^{\Ta} f(\a),
\label{eq:averaging}
\end{equation}
where $ f(\cdot)$ represents the auxiliary network, and $\Ta$ is the number of frames of the enrollment utterance.

The second network computes a time-frequency mask, $\m$, given the mixture signal, $\y$, and the speaker embedding vector, $\e$, as $ \m = g(\y,\e)$, 
where  $g(\cdot)$ represents the speech extraction network. The extracted speech signal, $\xh$, is then  obtained as $\xh = \m \odot \y $, where $\odot$ is the element-wise multiplication. 

There are several approaches to combine the two networks using, e.g., feature concatenation, addition, or multiplication. We employ here the multiplicative combination~\cite{Samarakoon+2016}, where the output of the first hidden layer of the speech extraction network $g(\cdot)$ is multiplied with the speaker embedding vector $\e$ as proposed in~\cite{delcroixIcassp19}.
We jointly train both networks to minimize the mean square error (MSE) between the estimated and reference target speech.

\section{Neural activity driven speech extraction}
\label{sec:ADEnet}
SpeakerBeam has demonstrated high speech extraction performance. However, it requires a pre-recorded enrollment utterance that may not always be available. Moreover, there may be a mismatch between the enrollment utterance and the target speech in the mixture due to, e.g., different recording conditions. 

Here, we assume that we can have access to the speaker activity information instead of the pre-recorded utterance. Let $\p \in \{0,1\}$ be a signal representing the activity of a speaker. $\p$ takes a value of $1$ when the speaker is speaking and $0$ otherwise. We consider two cases, i.e. (1) speaker activity \emph{with overlap}, where $\p$ contains speaker active regions, including regions where the interference speakers overlap with the target speaker, and (2) speaker activity \emph{without overlap}, where we remove the overlap regions from the speaker active regions based on the activity information of the interference speakers.

We first describe three configurations of ADEnet to exploit the speaker activity for target speech extraction shown in Fig.~\ref{fig:methods}(b),(c),(d). We then discuss how to obtain target speaker activity in practice. Finally, we discuss a training strategy to make ADEnet robust to estimation errors in the speaker activity signals.

\subsection{Three configurations of the proposed ADEnet}

The first configuration, shown in Fig.~\ref{fig:methods}-(b), exploits a similar architecture as SpeakerBeam, but replaces the enrollment utterance by the input mixture and the time-averaging operation of Eq.~(\ref{eq:averaging}) by a weighted-sum as,
\begin{equation}
\setlength{\abovedisplayskip}{1pt}
\setlength{\belowdisplayskip}{1pt}
    \e = \frac{1}{\sum_t \p} \sum_t \p f(\y) .
    \label{eq:weighted_sum}
\end{equation}
This formulation is similar to SpeakerBeam using instead of pre-recorded enrollment utterance the regions of the mixture signal where the target speaker is active. If we can remove the overlapping regions based on the speaker activity information (i.e., activity without overlap), the enrollment utterance consists of the regions where the target speaker is the single active speaker. This naturally avoids any mismatch in the recording conditions between the enrollment and the mixture. We call this approach \emph{ADEnet-auxiliary}.

An alternative option consists of simply concatenating the speaker activity $\p$ and the speech mixture $\y$ at the input of the network. This approach does not compute any explicit speaker embedding vector but expects that the extraction network can learn to track and identify the target speaker internally based on its activity. Consequently, it does not use any auxiliary network. This approach is shown in Fig. \ref{fig:methods}-(c). We call it \emph{ADEnet-input}.

The third configuration is shown in Fig. \ref{fig:methods}-(d). It consists of a combination of both approaches, where we use the speaker activity at the input and also to compute the speaker embedding vector with Eq. (\ref{eq:weighted_sum}). We call this approach \emph{ADEnet-mix}.

\subsection{Speaker activity information}
ADEnet requires the speech activity of the target speaker in the mixture. We aim at building a versatile system that can perform speech extraction independently of the way we obtain the speaker activity.

In practice, there are several possible options to obtain the speaker activity. Naturally, a conventional VAD~\cite{webrtcvad}, would not work in this context because it could not discriminate the target speaker from the interference in a mixture. 
However, if we have access to video recordings of the target speaker, we could use a visual-based VAD to detect the speaker activity~\cite{sodoyer2006analysis,liu2004voice}.
If enrollment utterances are available, we can obtain the speaker activity  with a personal VAD~\cite{Ding2020}.
If none of the above are available, we can use diarization~\cite{Medennikov2020TargetSpeakerVA,GarciaRomero2017SpeakerDU,Fujita2020EndtoEndND} to obtain the speech activity of the speakers in a recording.

As an example use-case, we investigate using ADEnet in a meeting scenario. We use diarization to obtain the speech activity of all speakers in a meeting. We then apply ADEnet to each speaker identified by the diarization system. This builds upon the great progress of recent diarization systems that can achieve a low DER in meeting-like situations~\cite{Fujita2020EndtoEndND,Medennikov2020TargetSpeakerVA,Huang2020SpeakerDW}. Combining diarization with ADEnet offers a practical yet relatively simple approach for enhancing ASR performance in overlapping conditions. 

However good recent VAD or diarization systems are, we cannot expect to obtain oracle speaker activity information. Besides, depending on the scenario, it may not be possible to obtain speaker activity without overlap. It is thus essential to develop a system robust to errors in the speaker activity detection.
In the next section, we explain a training strategy making ADEnet robust to errors in the activity signals.

\subsection{Training of AEDnet}
\label{ssec:training}
We require the triplet of the observed mixture $\y$, the reference target speech $\x$, and the target speaker activity $\p$ to train ADEnet. We use simulated speech mixtures and extract the \emph{oracle speaker activity} by applying a conventional VAD approach~\cite{webrtcvad} on the reference target speech. Besides, we use speaker activity without overlap, to allow the system to better capture the characteristics of the target speaker. 

Using oracle speaker activity information obtained from the reference signals allows us to build a system independent of the method used for estimating the speaker activity at test time. However, it is unrealistic to assume the availability of oracle error-free speaker activity information. Therefore, we include noise to the speaker activity information at training time to make the system robust. 
Practically, for each oracle speech segment found by the VAD, we modify the segment boundaries by adding to the start and end times a value uniformly sampled between $-1$ and $1$ seconds.

\section{Related work}
\label{sec:related}
There have been several works that integrated speech activity into non-neural multi-channel speech separation systems to extract speech.~\cite{rivet2007visual,Ono12,Nesta17,Bddeker2018FrontendPF}. 
For example, supervised independent vector analysis (IVA) modifies the objective function of IVA  by including an extra-term related to the target speaker, such as the source activity, allowing the system to extract that speaker~\cite{Nesta17}.

The diarization information has also been combined with spatial clustering-based separation in the guided source separation (GSS) framework~\cite{Bddeker2018FrontendPF}. GSS introduces the speaker activity as a prior for the mask estimation of the complex angular central Gaussian mixture model (cACGMM) based source separation, enabling identifying the target source from the cACGMM outputs.
GSS has been  applied successfully to severe recording conditions~\cite{Bddeker2018FrontendPF}.
Note that supervised IVA and GSS require multi-microphone recordings, while our proposed ADEnet can work with a single microphone. 

If multi-microphone recordings are available, we could extend  GSS with ADEnet by using time-frequency masks of the sources obtained from ADEnet instead of the speaker activity as priors to cACGMM as proposed in~\cite{nakatani_17} for speech separation. Such an extension will be part of our future works.

\section{Experiments}
\label{sec:expe}
We performed two types of experiments, with (1) simulated two-speaker mixtures to show the potential of using speaker activity for speech extraction and (2) meeting-like recordings to illustrate a potential use-case of ADEnet as a pre-processor for ASR. 

\subsection{Dataset}
We used the same training data for all experiments. It consisted of 63000 mixtures of reverberant two-speaker mixtures with background noise at SNR between 10 and 20 dB. The speech signals were taken from the LibriSpeech corpus\cite{Panayotov2015librispeech}. 
We further created a validation and test set with the same conditions. The test set consisted of 1364 mixtures with an averaged overlapping ratio of 38.5~\%.

In the second experiment, we used the meeting-like LibriCSS corpus~\cite{Chen2020ContinuousSS}, which consists of 8-speaker meeting-like recordings sessions of 10 minutes, obtained by re-recording LibriSpeech utterances played through loudspeakers in a meeting room. The overlap ratio varies from 0 to 40 \%. 

\subsection{Auxiliary information}
For training, we used the ``oracle'' speaker activity without overlap obtained by applying the webrtc VAD~\cite{webrtcvad} to the reference signals. 
We trained the system with or without \emph{noisy activity training} (adding noise to the speaker activity signals, as described in \ref{ssec:training}). 
In the first experiment with simulated two-speaker mixtures, we used the oracle speaker activity with and without noise. In the second experiment with LibriCSS, we used target speaker VAD (TS-VAD) based diarization~\cite{Medennikov2020TargetSpeakerVA} to obtain the speech activity regions of each speaker.

\subsection{Experimental settings}
\label{ssec:settings}
In all experiments, we used a similar architecture for the extraction network, which consisted of 3 bidirectional long short-term memory (BLSTM) layers with 1200 units, and two fully connected (FC) layers with rectified linear unit (ReLU) activation functions. The auxiliary network consisted of two FC layers with 64 hidden units and ReLU activation functions. We used the same configuration for the SpeakerBeam network. The input features consisted of 257-dimensional log amplitude spectrum coefficients obtained using a short-time Fourier transform with a window of 32 msec and a shift of 8 msec. The training loss consisted of the MSE between the spectra of the reference signal and the extracted speech. 
We trained the networks using the ADAM optimizer using the padertorch toolkit~\cite{padertorch}.

\subsection{Experiments on simulated two-speaker mixtures}

\begin{table}[tb]
  \caption{Speech enhancement results in terms of SDR [dB] for simulated 2-speaker data with ADEnet without and with noisy activity training. The SDR of the mixture signal is -0.2 dB, and the SDR of SpeakerBeam is 9.4 dB.}
\setlength{\tabcolsep}{3pt}
\renewcommand{\arraystretch}{0.8}
  \label{tab:results_sim}
  \centering
  \begin{tabular}{lc|c|c|c|c|c}
    \toprule
 & Noisy& \multicolumn{4}{c|}{\emph{Activity signal at test time}}& \\
& activity & \multicolumn{2}{c|}{w/o overlap} &  \multicolumn{2}{c|}{w/ overlap} \\
& training & Oracle & +Noise & Oracle & +Noise & Avg \\
\midrule
ADEnet-aux & -   &	9.7	&	6.0&6.0&	4.3 & 6.5\\
ADEnet-in	& -    &\bf{10.2}& 4.3&		3.7	&2.8  & 5.3\\
ADEnet-mix & -	&10.1&	3.9&	3.7&	3.0 & 5.2 \\
\midrule
ADEnet-aux	& \checkmark &	8.4	&8.5	&\bf{7.9}	&8.0	&8.2\\
ADEnet-in	& \checkmark &	9.0	&9.2	&7.7	&\bf{8.6}	&8.6\\
ADEnet-mix	& \checkmark &	9.2	&\bf{9.4}	&\bf{7.9}	&\bf{8.6}	&\bf{8.8}\\
\bottomrule
  \end{tabular}
\end{table}	

We first performed experiments on the simulated two-speaker mixtures. Table \ref{tab:results_sim} shows the signal to distortion ratio (SDR)~\cite{vincent2006performance} obtained with ADEnet with and without noisy activity training and for different activity signals at test time. At test time, we used oracle speaker activity signals without overlap as during training (i.e. ``w/o overlap'') without, and with adding noise, i.e. ``Oracle'' and ``+Noise'', respectively. We also explored training and test mismatch by using speaker activity with overlap at test time (i.e. ``w/ overlap'').  The SDR of the mixture signals is -0.2 dB. 
We compare the proposed ADEnet with SpeakerBeam using pre-recorded enrollment utterances. We used a randomly selected utterance of the target speaker from the LibriSpeech corpus as the enrollment utterance.  SpeakerBeam achieved an SDR of 9.4 dB.
Besides, masking out the regions where the speaker is inactive from the observed signal (``inactivity-masking'') leads to an SDR or 3.8 dB with oracle speaker activity, and -1.6 dB when using noisy speaker activity.

The upper part of Table \ref{tab:results_sim} shows the results without noisy activity training.
The first column confirms that ADEnet with oracle speaker activity can achieve higher SDR than SpeakerBeam (10.2 dB vs. 9.4 dB) without the need for pre-recorded enrollment utterance. 
Moreover, it greatly outperforms the simple ``inactivity-masking''. These results demonstrate that ADEnet can effectively extract speakers based on the speaker activity.
However, the SDR degrades significantly when precise speaker activity is not available as revealed by the poor results when including noise or overlap regions. 

\begin{figure}[tb]
\begin{minipage}[t]{1.0\linewidth}
  \centerline{\includegraphics[width=3.4cm]{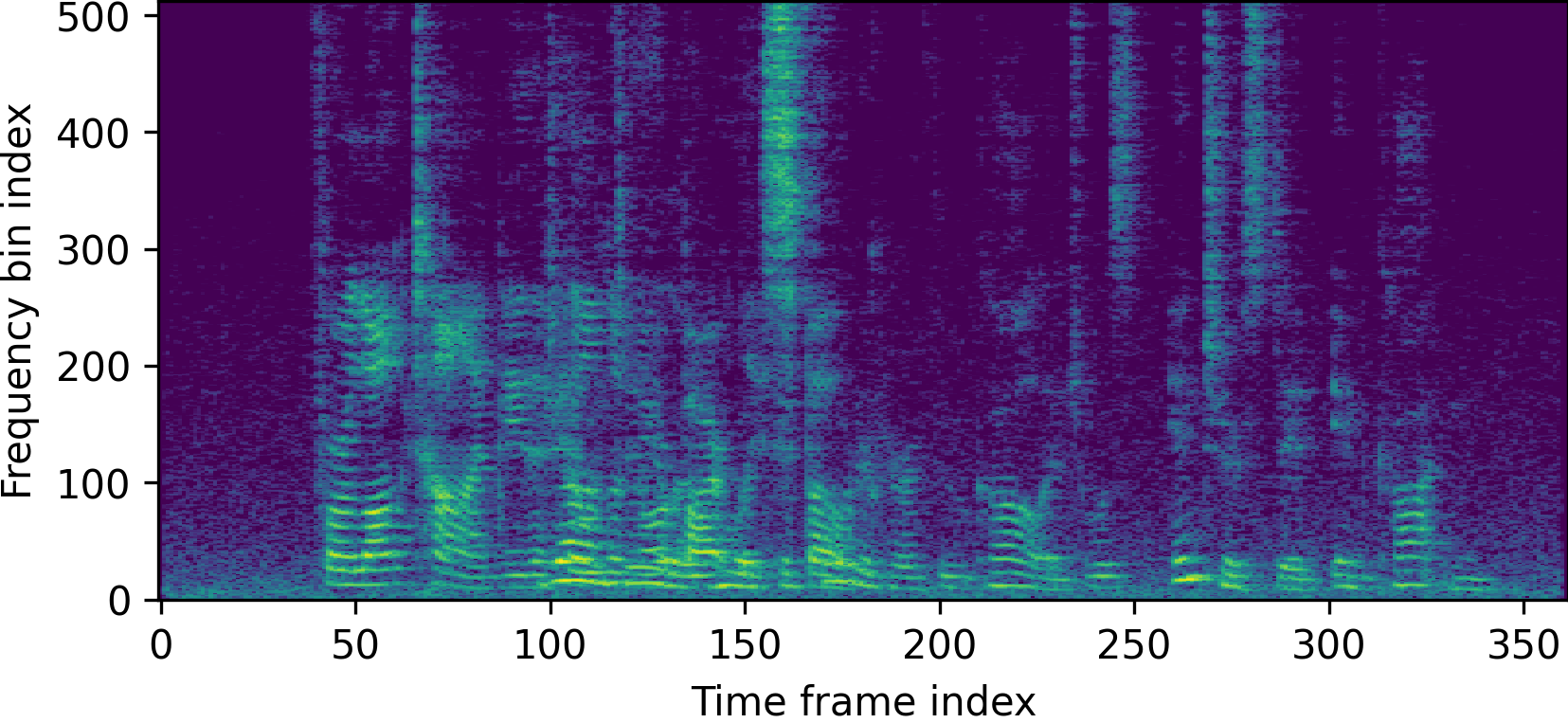}}
  \centerline{(a) Mixture}
\end{minipage}
\begin{minipage}[t]{0.48\linewidth}
  \centerline{\includegraphics[width=3.4cm]{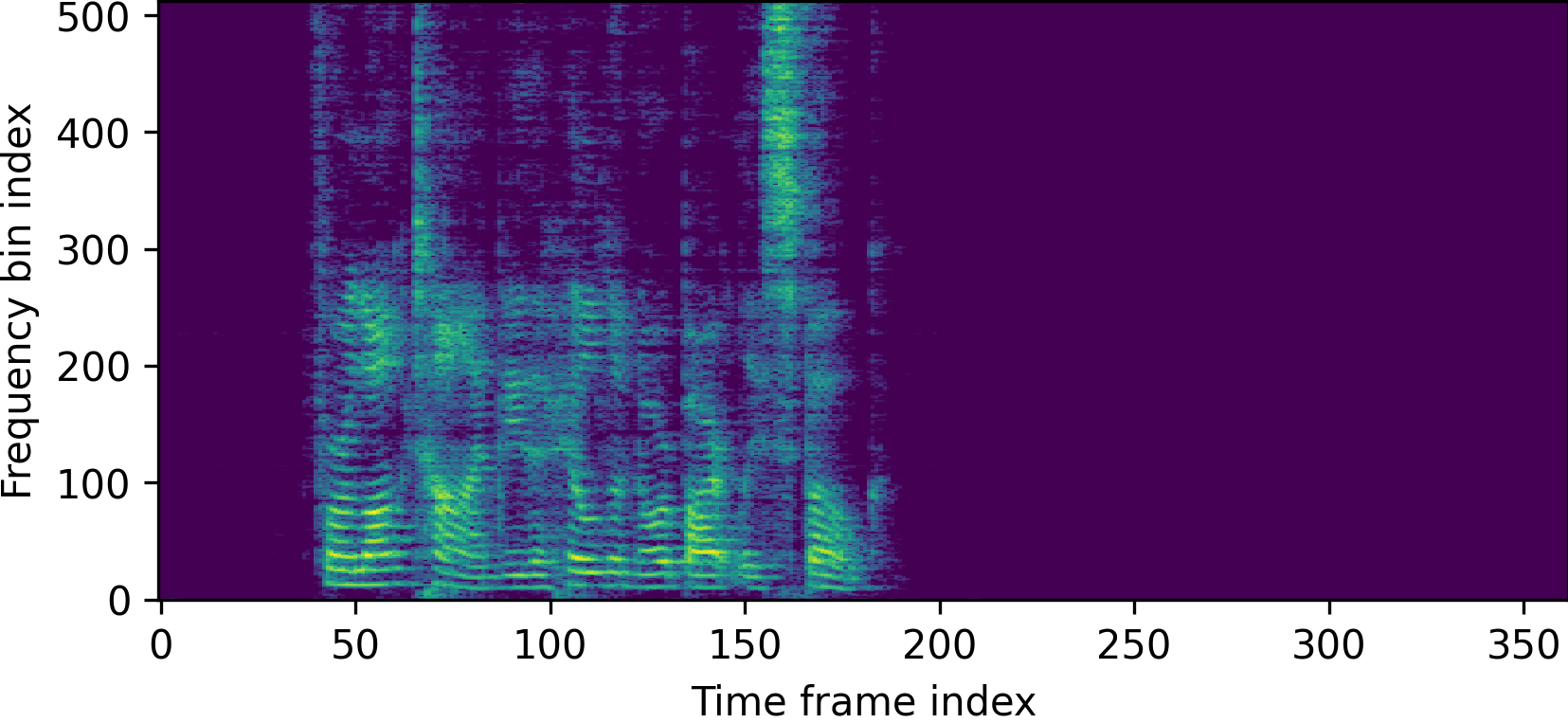}}
  \centerline{(b) Reference speaker 1}
  \centerline{\includegraphics[width=3.4cm]{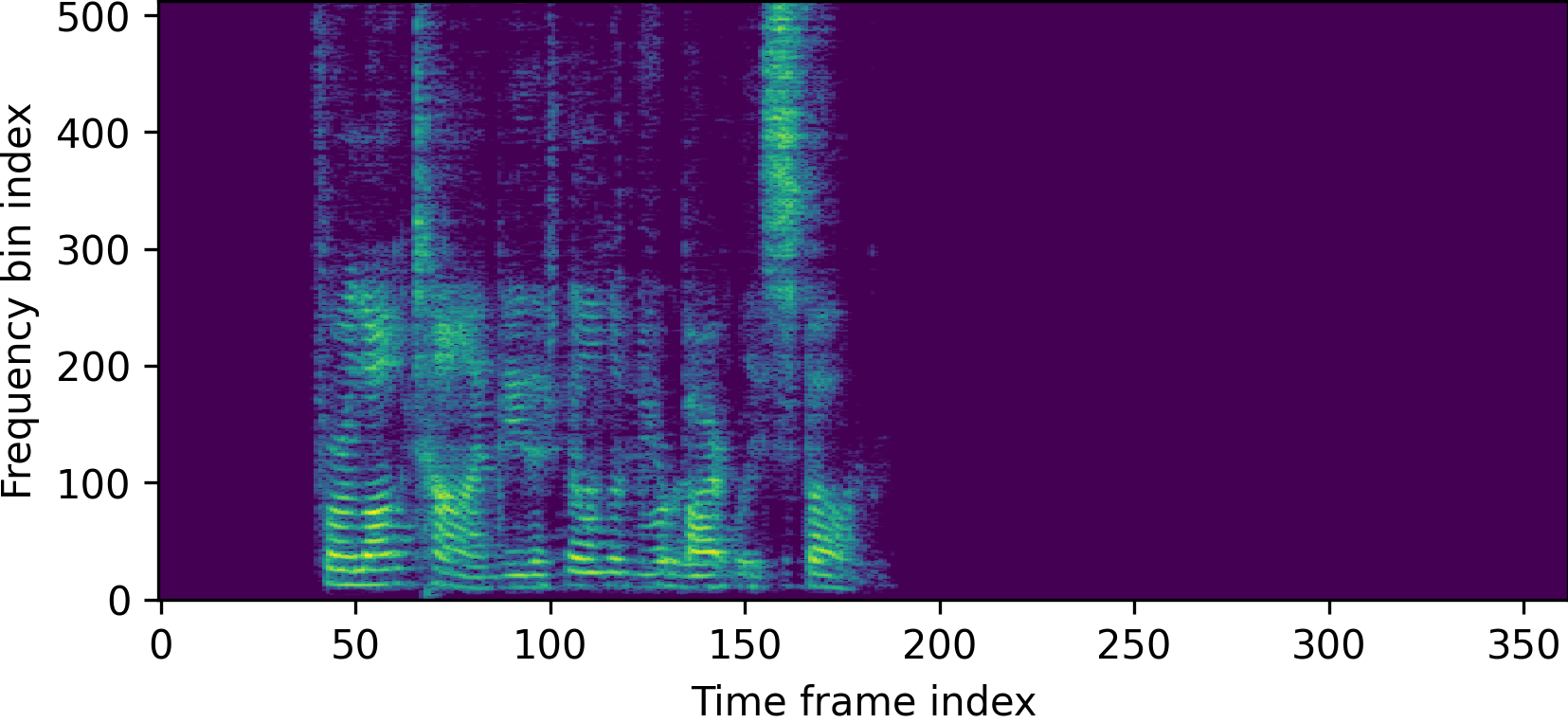}}
  \centerline{(d) Extracted speaker 1}
\end{minipage}
\begin{minipage}[t]{0.48\linewidth}
  \centerline{\includegraphics[width=3.4cm]{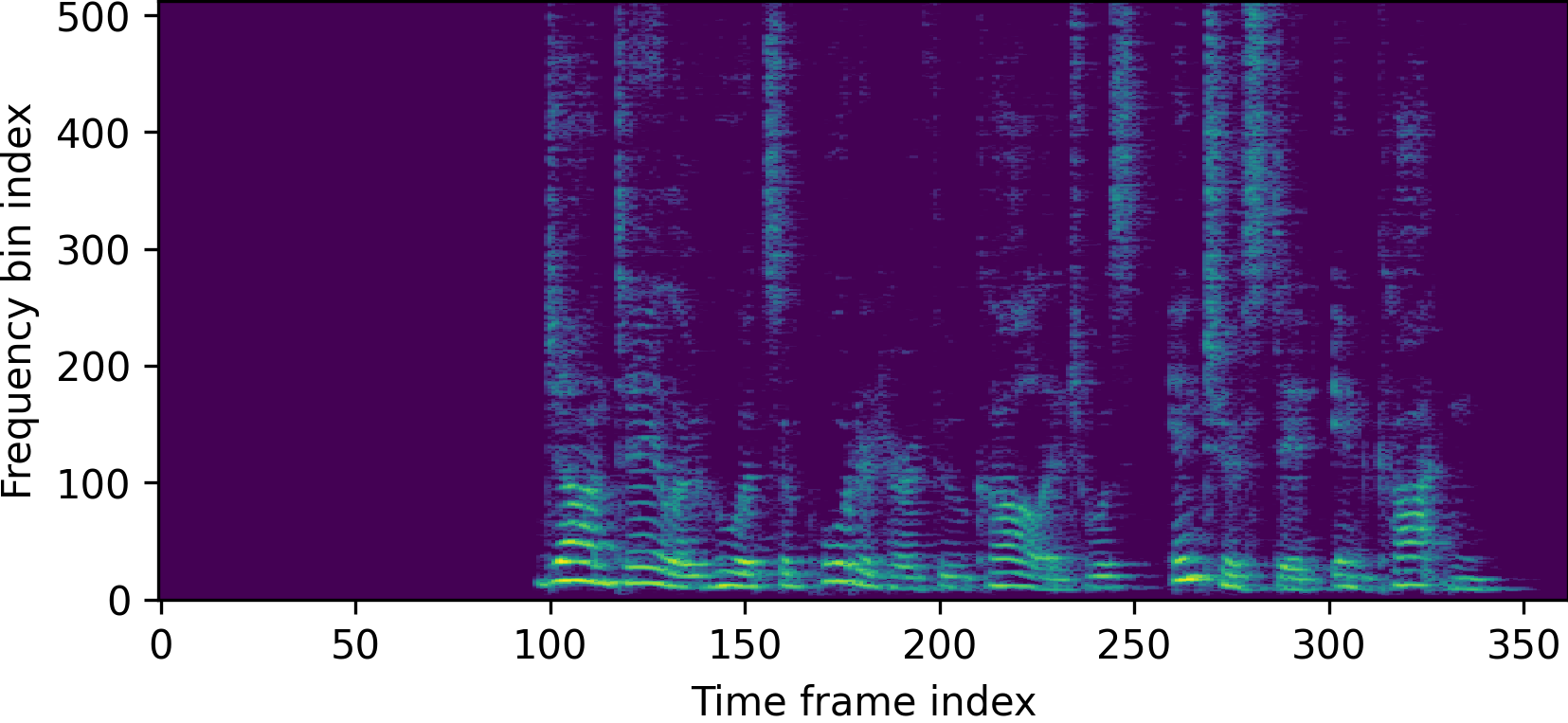}}
  \centerline{(c) Reference speaker 2}
  \centerline{\includegraphics[width=3.4cm]{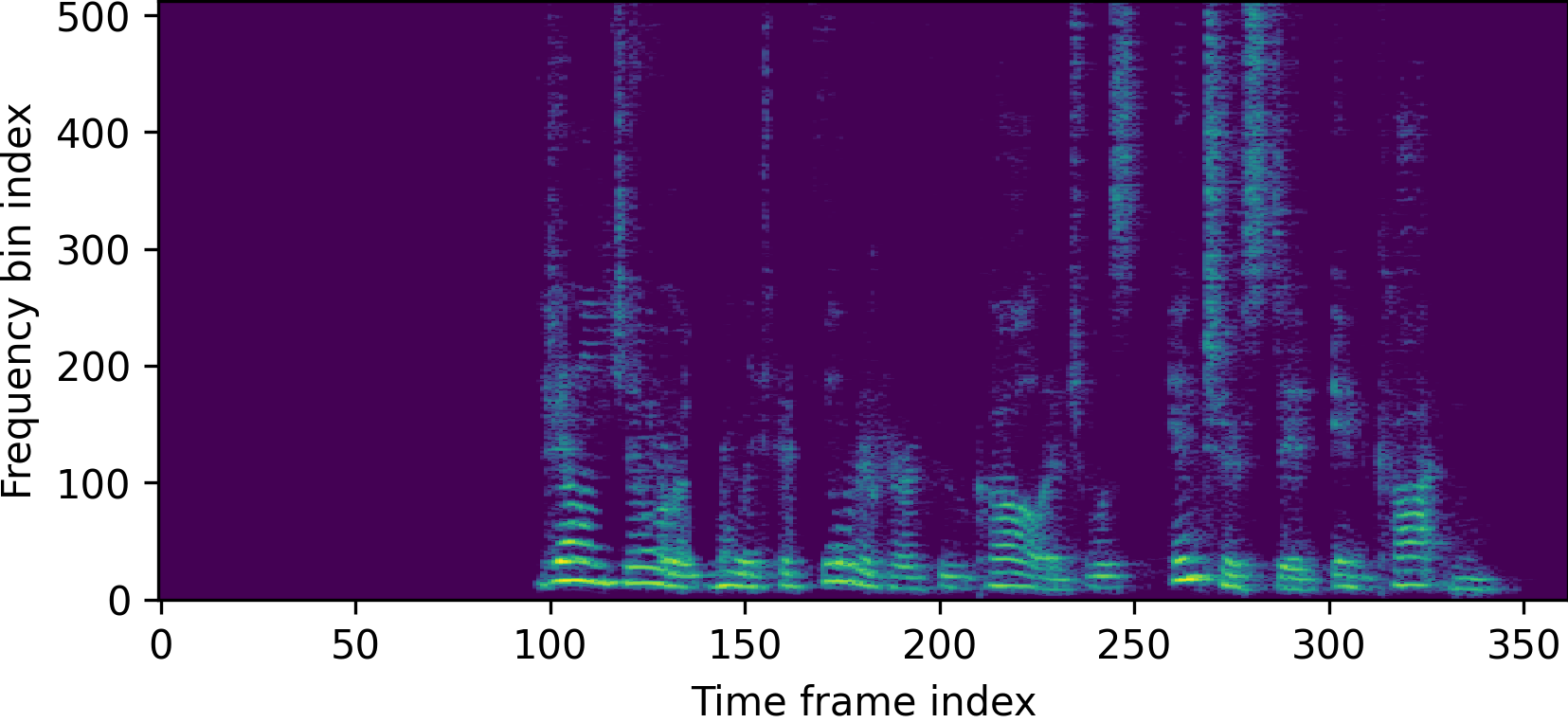}}
  \centerline{(e) Extracted speaker 2}
\end{minipage}
\vspace{-2mm}
\caption{Example of extracted speech with ADEnet-mix.}
\label{fig:res_enh}
\vspace{-0.4cm}
\end{figure}

The lower part of Table~\ref{tab:results_sim} shows the results with noisy activity training. The oracle performance degrades, but the systems become much more robust to errors in the activity signals. Overall, the performance in noisy conditions becomes a little worse than SpeakerBeam but still much better than using ``inactivity-masking''.
ADEnet-mix with noisy training achieves superior performance on average, and we use this configuration in the following experiments.
Figure \ref{fig:res_enh} shows an example of extracted speech, showing that ADEnet can extract a speaker even in the overlapping regions.

This experiment demonstrates that single-channel speech extraction is possible without explicit use of enrollment utterances or video. Moreover, the relatively stable performance with various quality of the activity signals indicates that ADEnet could be used with various diarization or VAD methods that provide or not speaker regions without overlap.

\subsection{Experiments on LibriCSS meeting-like data}
We present results on the LibriCSS meeting-like corpus, using the TS-VAD-based diarization to obtain the speech activity regions of all speakers in the meeting. 
We use TS-VAD as an example of diarization, but other diarization approaches could be used such as, e.g., ~\cite{GarciaRomero2017SpeakerDU,Huang2020SpeakerDW}.   TS-VAD achieved a DER of 7.3 \%~\cite{Desh_SLT21}. 
We use the ESPnet toolkit and the transformer-based recipe developed for LibriCSS~\cite{Watanabe2018espnet, espnet_libricss} as ASR back-end, which provides a strong baseline for the task~\cite{Desh_SLT21}. 

Diarization finds the speech activity for the speakers in the recording. 
From the diarization output, we process each continuous speech segment separately, by adding 2 seconds of context at the beginning and end to create extended segments. We used the speaker activity estimated by the diarization within the extended segment as speaker activity signal. Besides, we also experimented with speaker activity without overlap by removing  from the activity signal the regions where multiple speakers were detected. ASR is performed separately on each speech segments found by the diarization and evaluated using the concatenated minimum-permutation word error rate (cpWER)~\cite{Watanabe2020CHiME6CT} that includes the errors caused by diarization.

Table \ref{tab:asr} shows the cpWER obtained with the diarized output with and without ADEnet. Note that ADEnet operates on single-channel recordings.
Even if we used a powerful diarization, we see the degradation caused by the overlapping speech.
 We confirm that of ADEnet consistently improves cpWER, especially for the high overlapping regions, with relative WER improvement ranging from 6.5~\% to 25~\%. 
These results should be put into perspective with those obtained with a multi-channel system using GSS~\cite{Bddeker2018FrontendPF} with seven microphones, which achieved an average WER of 11.2 \%.
Compared to GSS, ADEnet has the advantage that it can work with single-channel recordings. The integration of ADEnet with GSS is possible and will be part of our future works.

\begin{table}[tb]
	\caption{cpWER using TS-VAD-based diarization, the proposed ADEnet and a Transformer-based ASR backend. 
	}
	\label{tab:asr}
\setlength{\tabcolsep}{2pt}
	\centering
    \begin{tabular}{lccccccccc}
    \toprule
   & w/o & \multicolumn{7}{c}{Overlap ratio in  \% } \\ 
   & overlap     & 0L &	0S	& OV10	& OV20	& OV30	& OV40	& Avg  \\ 
    \midrule 
    no proc & na & 11.2 &	9.4 & 16.2 & 23.1 &	33.6 &	41.1 &	23.9 \\
    ADEnet  & - &  \bf{10.5} &	\bf{8.5} & 13.8 & 18.7 &	26.5 &	30.7 &	19.2 \\
    ADEnet & \checkmark &\bf{10.5}&	8.6	&\bf{13.6}&	\bf{18.3} &	\bf{25.8}	&\bf{29.8}	&	\bf{18.8} \\
    \bottomrule
    \end{tabular}
\end{table}

\section{Conclusion}
\label{sec:conclusion}
We investigated the potential of using speaker activity to develop a simple and versatile neural speech extraction. We showed experimentally that the proposed ADEnet with precise speaker activity information could achieve high single-channel speech extraction performance. Besides, combined with diarization, it can significantly improve the ASR performance in meeting-like situations.

For further improvements, we plan to investigate more powerful network configurations~\cite{delcroix2020improving} and combine ADEnet with multi-microphone schemes~\cite{Bddeker2018FrontendPF,nakatani_17}. Besides, we would like to compare ADEnet with visual clue based speech extraction approaches.

\small
\textbf{Acknowledgment.} This work was started at JSALT 2020 at JHU, with support from Microsoft, Amazon, and Google. We thank Desh Raj, Pavel Denisov, and Shinji Watanabe for providing the LibriCSS baseline.

\bibliographystyle{IEEEbib}
\bibliography{refs}

\end{document}